\documentclass{aa}  

\usepackage{graphicx}
\usepackage{txfonts}
\usepackage{xcolor} 
\usepackage{soul}
\definecolor{mygreen}{RGB}{0, 185, 118}

\begin{document}

   \title{Helmet streamer influence on the evolution of magnetic flux ropes }

   \author{M. Cécere \inst{1,2}
          \and
          P. F. Wyper\inst{3}
          \and
          G. Krause\inst{4}
          \and
           A. Sahade\inst{5}
           \and
           O. E. K. Rice\inst{3}
          }

          \institute{
          Instituto de Astronom\'{\i}a Te\'orica y Experimental, CONICET-UNC, C\'ordoba, Argentina.\\
          \email{mariana.cecere@unc.edu.ar}
          \and
          Observatorio Astron\'omico de C\'ordoba, UNC, C\'ordoba, Argentina.
          \and
          Department of Mathematical Sciences, Durham University, Durham DH1 3LE, UK.
          \and
          Instituto de Estudios Avanzados en Ingenier\'{\i}a y Tecnolog\'{\i}a, CONICET-UNC, C\'ordoba, Argentina.
          \and
          NASA Goddard Space Flight Center, Greenbelt, MD 20771, USA.\\  
             }

   \date{}%Received September 15, 1996; accepted March 16, 1997}

% \abstract{}{}{}{}{} 
% 5 {} token are mandatory
 
  \abstract
   {  
   Solar eruptions are of great interest for space weather. Understanding the physical mechanisms that influence their evolution is essential for improving future predictions about the geoeffectiveness of the event. Whether during periods of minimum or maximum solar activity, helmet streamers (HSs) are present in the solar corona. This magnetic structure features a current sheet 
   of low magnetic energy, where coronal mass ejection events tend to deflect toward. However, it also includes a closed magnetic field region
   underlying this current sheet, 
   where eruptions are often confined. This makes it an interesting structure to study, as the inherent complexity of the structure hinders the predictability of the eruption.
   }
   {The aim of this study is to investigate the influence of the HS on the evolution and potential confinement of a magnetic flux rope (MFR). The idea is to explore magnetic configurations involving the MFR and HS that are more likely to allow the MFR to rise through the overlying magnetic field, with the ultimate goal of establishing simple parameters that can help predict the conditions under which an MFR may ascend or remain confined.}
   {Through 2.5D MHD numerical simulations, we emulate the dynamics of MFRs in the presence of a HS. We analyse the dynamics of different magnetic field configurations, paying special attention to the mechanisms that facilitate the ascent or confinement of the MFR.   }
   {We find that the null point reconnection mechanism plays a fundamental role in the dynamics of the MFR. Depending on the initial configuration, null point reconnection can either confine the ascent by disrupting the MFR or facilitate its rise by reducing the strapping flux above it.  
   We also identify a critical value in the relationship between the magnetic flux that the MFR must traverse during its ascent and its own magnetic flux.
   We find that if the strapping flux above
   the MFR is less than two-thirds of its own poloidal flux, the MFR is able to ascend successfully. }
   {We conclude that in our simulations null point reconnection plays a major role in facilitating the ascent of the MFR. A key factor in predicting whether the MFR will rise successfully is the initial ratio between its poloidal flux and the strapping magnetic flux above it.   
   }

   \keywords{Sun: coronal mass ejections (CMEs) -- Sun: magnetic fields --
                Magnetohydrodynamics (MHD) --
                Methods: numerical
               }

   \maketitle
%
%-------------------------------------------------------------------

\section{Introduction}
CMEs play a crucial role in space weather, and predicting their occurrence requires understanding not only the trajectories but also the likelihood of an event. In this context, several studies analyse the physical nature of the magnetic flux rope (MFR) eruption through the study of magnetic fluxes. For example, \cite{2020ApJ...898L..12Z} demonstrate that an increase in magnetic flux, primarily axial, within the MFR drives the rope to evolve toward a critical condition, eventually triggering its eruption, indicating a threshold for the eruption to occur. On the other hand, using the catastrophe model \citep{2000JGR...10523153F}, it was shown that while ejection is related to an increase in magnetic flux, a decrease in this quantity also supports the ejection \citep{2018JGRA..123.2513Z}.

Other studies emphasise the crucial role of the ambient environment in determining ejection outcomes. In the context of flux rope eruptions \citet{2023ApJ...955..114R} showed that there exists a clear threshold in ratios between the magnetic flux in a flux rope and various quantities influenced by the strength of the background field, above which an eruption is very likely. They also tested the predictability of these eruptions against measures such as the eruptivity index \citep{Pariat2017}, finding that the large peaks originally observed in the eruptivity index prior to eruptions were due to the orientation of the ambient magnetic field. 
In the context of pseudostreamer eruptions \cite{2022A&A...663A..32T} demonstrated that neighbouring magnetic structures are important because they can decelerate the eruption through magnetic reconnection, creating overlapping arcs that increase confinement. Moreover, this interchange (open-closed) reconnection leads to the opening of a considerable fraction of the MFR flux, partially altering the MFR's path and affecting the CME deflection \citep{2023ApJ...957...74B}. 

\cite{2013ApJ...771...82M} demonstrated in a null point configuration beneath a helmet streamer (HS) that when the amount of overlying closed flux is small compared to the erupting flux, such as when a coronal hole is near the eruption, interchange reconnection is likely to occur. Conversely, if the amount of overlying closed flux is large, interchange reconnection becomes highly unlikely. Similar findings were reported in \cite{2022A&A...662A.113S}, where the eruption occurred inside a pseudostreamer structure. This study revealed that significant overlying/magnetic cage flux quantities act as a preventive factor, hindering the eruption. Similarly, \citet{Karpen2024} recently reported on an observed confined eruption within a pseudostreamer, comparing it to a 3D simulation, where again significant overlying flux played an important role.

The goal of this paper is to analyse the evolution of MFR fluxes through their interaction with the surrounding magnetic field fluxes. While this study focuses on understanding the factors that influence the ascent and confinement of MFRs, the ultimate objective is to contribute to future efforts in predicting their potential eruptivity. We assume an unstable magnetic flux rope superimposed with a HS structure (Section~\ref{s:model}). Null point reconnection is shown to play both a stabilising and destabilising role on the evolution, depending upon the configuration. We show in some cases the flux rope is destroyed or reaches a new equilibrium. In these stable cases there is significant overlying flux that also acts to suppress the eruption (Section~\ref{s:results}). We demonstrate that overlying flux
erosion by null point reconnection plays a key role in achieving a successful ascent (in the manner of \citet{2013ApJ...771...82M}), with the caveat that there is no solar wind in our simulation. However, the direction of this reconnection is key, and the opposite scenario of the flux rope tunnelling deeper into the closed field can also occur. This  is
followed by a discussion and our conclusions in Section \ref{s:conclus}.

%--------------------------------------------------------------------
\section{Model}
\label{s:model}
To study the evolution of a MFR embedded in a HS background, we numerically solve the 2.5D ideal MHD equations, taking into account a stratified atmosphere. The equations in their Cartesian conservative form with CGS units are expressed as follows:
\begin{equation}\label{e:cont}
\frac{\partial\rho}{\partial t}+\nabla\cdot(\rho\vec{v})=0 \, ,
\end{equation}
\begin{equation}\label{e:euler}
\frac{\partial (\rho \vec{v})}{\partial t} + \nabla \cdot \left(\rho \vec{v} \vec{v} - \frac{1}{4\pi} \vec{B}\vec{B} \right) + \nabla p + \nabla\left( \frac{B^2}{8\pi}\right)   = \rho \vec{g} \, ,
\end{equation}
\begin{equation}\label{e:consE}
\frac{\partial E}{\partial t} + \nabla \cdot \left[\left(E + p + \frac{B^2}{8\pi}\right)\vec{v} -\frac{1}{4\pi} \left(\vec{v\cdot B}\right)\vec{B}\right] = \rho \vec{g}\cdot\vec{v} \, ,
\end{equation}
\begin{equation}\label{e:induccion}
\frac{\partial \vec{B}}{\partial t} + \nabla \cdot \left(\vec{v} \vec{B} - \vec{B} \vec{v} \right) = \vec{0} \, ,
\end{equation}
where $\rho$ represents the plasma density, $p$ the gas pressure, $\vec{v}$ the velocity, $\vec{B}$ the magnetic field, and $\vec{g}$ the acceleration due to gravity. $E$ is the total energy (per unit volume), given by  
\begin{equation*}
    E = \rho \epsilon + \frac{1}{2} \rho v^2 + \frac{B^2}{8\pi},
\end{equation*}
where $\epsilon$ is the internal energy. In this system, the current density is $\vec{j}=\frac{c}{4\pi}{\nabla\times}\vec{B}$ (with $c$ being the speed of light) and the magnetic field must satisfy the divergence-free condition $ \nabla\cdot \vec{B} = 0$. 
We consider a fully ionised hydrogen plasma as the medium, where the perfect gas law applies: $p = \rho R T/\mu = (\gamma-1)\rho\epsilon$. Here, $R$ represents the gas constant, $T$ is the plasma temperature, $\mu$ stands for molar mass, and $\gamma$ is the specific heat ratio, equal to $5/3$.

We performed simulations using the FLASH Code \citep{2000ApJS..131..273F}, specifically the fourth version, which is equipped with the unsplit staggered mesh solver. This solver employs a second-order directionally unsplit scheme with a MUSCL-type reconstruction and a constrained-transport method to satisfy the divergence-free condition of the magnetic field \citep{2009JCoPh.228..952L}.
We use an atmosphere that includes a thin chromosphere to avoid cavitation effects brought on by the rapid expansion of the flux ropes in our simulations. %In addition, 
As a consequence of the presence of the chromosphere, a strong stratification is produced at the base of the atmosphere. In this region the thermodynamic variables change by two orders of magnitude over a relatively short distance, which is covered by a few grid cells even when the maximum refinement level is used. 
This condition together with the need to perform long time simulations cause MUSCL methods with conventional reconstruction schemes to fail due to the emergence of spurious momentum fluxes in the gravity direction that artificially perturb the background hydrostatic equilibrium.
For this reason, we implement the well-balanced scheme proposed by \cite{2019A&A...631A..68K}, which uses a hydrostatic reconstruction scheme to ensure the numerical equilibrium for constant and linear temperature distributions in equilibrium atmospheres.
Regarding the boundary conditions, we proceed as follows:
at the lateral boundaries, we applied an outflow condition for thermodynamic variables and a linear extrapolation for the magnetic fields to maintain the initial force-free configuration. For the upper and lower boundaries, we implemented a hydrostatic condition with constant temperature extrapolation \citep{2019A&A...631A..68K}, and at the lower boundary, we also adopted a line-tied condition. Neglecting magnetic resistivity, we opted for the ideal MHD equations, capitalising on the numerical diffusion in the simulations for the required dissipation, as detailed in \cite{2018MNRAS.474..770K}. The simulation's highest resolution corresponds to cells of approximately $[0.8\times0.8]\,\textrm{Mm}^2$ 
within a physical domain of $[-2000,2000]\,\textrm{Mm} \times [0,8000]\,\textrm{Mm}$, where the pressure and temperature gradients satisfied the refinement criterion. This simplification significantly diminished computational costs.

\subsection{Magnetic field configurations} 
\label{sec:magfield}
We adopt the catastrophe model by \cite{1990JGR....9511919F} as the fundamental structure for the magnetic flux rope in the simulation. In this scenario, an out-of-equilibrium magnetic configuration leading to the ascent of a MFR is established when a current-carrying wire, its image, and a magnetic dipole are present. We also introduce a current distribution solely in the $\phi$ direction within the MFR to generate a helical magnetic field and lower the equilibrium gas pressure. This allows us to express the magnetic field of the MFR as follows:
\begin{align}\label{e:BFR}
      B_{x,\mathrm{MFR}}=-&B_\phi(R_-)\tfrac{(y-h_0)}{R_-} +  B_\phi(R_+)\tfrac{(y+h_0)}{R_+} - \nonumber \\ &MdB_\phi{\scriptstyle\left(r+\tfrac{\Delta}{2}\right)}\left(r+\tfrac{\Delta}{2}\right)\tfrac{(x-x_{\mathrm{MFR}})^2-(y+d)^2}{R_d^4} \, ,\nonumber \\
      B_{y,\mathrm{MFR}}=\quad &B_\phi(R_-)\tfrac{x-x_{\mathrm{MFR}}}{R_-} - B_\phi(R_+)\tfrac{x-x_{\mathrm{MFR}}}{R_+}-\nonumber \\
      & MdB_\phi{\scriptstyle\left(r+\tfrac{\Delta}{2}\right)}\left(r+\tfrac{\Delta}{2}\right)\tfrac{2(x-x_{\mathrm{MFR}})(y+d)}{R_d^4}  \, , \nonumber \\
      B_{z,\mathrm{MFR}}=\quad &B_{\text{z}}(R_-)\, .
\end{align}
Here, $h_0$ represents the initial height of the MFR, $x_{\mathrm{MFR}}$ its position in $x-$axis, $M$ is the strength of the line dipole at depth $d$, $r$ is the radius of the current wire, $\Delta$ denotes the thickness of the transition layer between the current wire and the exterior. Additionally, $R_\pm= \sqrt{(x-x_{\mathrm{MFR}})^2+(y\pm h_0)^2}$ and $R_d= \sqrt{(x-x_{\mathrm{MFR}})^2+(y+d)^2}$ correspond to the distances measured from distinct reference points, specifically the image ($+$) and current ($-$) wire for $R_\pm$, and the dipole for $R_d$. Also, in terms of $j_0$, the current density of the current wire, the $\phi$-component of the magnetic field is,
\begin{equation} \label{e:Bphi}
    B_\phi{(R)}\!=\!
    \left\{
     \begin{array}{rl}
     \begin{alignedat}{2}
       &\tfrac{2\pi}{c}j_0R  && 0\leq R < r-\frac{\Delta}{2}\\
       &\tfrac{2\pi j_0}{cR}\left\{\tfrac{1}{2}\left(r-\tfrac{\Delta}{2}\right)^2-\left(\tfrac{\Delta}{\pi}\right)^2 +\right. 
          \\
       &\tfrac{R^2}{2}+\tfrac{\Delta R}{\pi}\text{sin}\left[\tfrac{\pi}{\Delta}\left(R-r+\tfrac{\Delta}{2}\right)\right]+ \qquad && r-\frac{\Delta}{2}\!\leq \!R\! <r+\frac{\Delta}{2}
          \\
      &\left.\!\!\!\left(\tfrac{\Delta }{\pi}\right)^2\cos\left[\tfrac{\pi}{\Delta}\left(R-r+\tfrac{\Delta}{2}\right)\right]\right\} \\
      &\tfrac{2\pi j_0}{cR}\left[r^2+\left(\tfrac{\Delta}{2}\right)^2-2\left(\tfrac{\Delta}{\pi}\right)^2\right]  && r+\frac{\Delta}{2} \leq R\,.
  \end{alignedat}
 \end{array} \right.
\end{equation}
Finally, the $z$-component of the magnetic field and its current distribution $j_{\phi}$ inside the MFR (null otherwise), are described by:
\begin{eqnarray}
      B_\mathrm{z}(R) = \tfrac{4\pi j_1}{c}\sqrt{\left(r-\tfrac{\Delta}{2}\right)^2-R^2} 
         & 0\leq R < r-\frac{\Delta}{2}\,, \label{e:Bfieldz} \\
      j_\phi(R) = j_1R\left[\sqrt{\left(r-\tfrac{\Delta}{2}\right)^2-R^2}\right]^{-1}
        & 0\leq R < r-\frac{\Delta}{2} \label{e:jphi}
\end{eqnarray}
where $j_1$ is a current density. To create a non-equilibrium configuration for the MFR, we selected the following parameters: $h_0=100\,\mathrm{Mm}$, $M=5$, $d=0\,\mathrm{Mm}$, $r=10\,\mathrm{Mm}$, $\Delta=1\,\mathrm{Mm}$, $j_0=40\,\mathrm{statA\,cm}^{-2}$, and $j_1=33.6\,\mathrm{statA\,cm}^{-2}$.
   
To this MFR configuration, we add the HS magnetic structure, which we assumed to be symmetrical relative to the $y$-axis, with a neutral point at $(0,y_N)$. Below the neutral point there is a closed arcade and open fields at the flanks with a neutral electric current sheet that stretches along the $y$-axis from $y=y_N$ to infinity. The magnetic field is potential everywhere except on the current sheet \citep{2001SoPh..200..115H}: 

\begin{equation}
     B_{x,\mathrm{HS}} - i B_{y,\mathrm{HS}} = B_0 \frac{(\omega + i\,y_N)^{1/2} (\omega - i\,y_N)^{1/2}}{F(a,b,y_N)} \ln{\left(\frac{\omega^2-a^2}{\omega^2-b^2}\right)}
\end{equation}
where $B_0$ is the strength of the magnetic field, $\omega=x+i\,y$, $a=0.5\times10^{3}\,\mathrm{Mm}$, $b= 1.5\times10^{3}\,\mathrm{Mm}$, $y_N=1.75\times10^{3}\,\mathrm{Mm}$, and
\begin{equation}
   \begin{aligned}
       F(a,b,y_N) =& \frac{1}{2(b-a)}\left[b(b^2+y_N^2)^{1/2} - a(a^2+y_N^2)^{1/2} \right. \\
                   & \left. + y_N^2 \ln{\left(\frac{b+(b^2+y_N^2)^{1/2}}{a+(a^2+y_N^2)^{1/2}}\right)}\right].
   \end{aligned}
\end{equation}

Then, we obtain the total magnetic field as follows:
\begin{align*}
    B_x&=B_{x,\mathrm{MFR}}+B_{x,\mathrm{HS}} \, ,\nonumber \\
    B_y&=B_{y,\mathrm{MFR}}+B_{y,\mathrm{HS}} \, , \nonumber \\
    B_z&=B_{z,\mathrm{MFR}}\, .
\end{align*}

It is worth noting that the combination of both structures does not guarantee the ascent of the MFR, as the HS field can influence its initial dynamics and potentially confine the structure, preventing the ascent.

\subsection{Thermodynamic variables}
We simulate a stratified atmosphere by using a multi-layer structure \citep{2012MNRAS.425.2824M}. The chromosphere extends from $y=0$ to $y=h_{\mathrm{ch}}=10\,\mathrm{Mm}$ with a uniform temperature $T_{\mathrm{ch}}=10^4\,\mathrm{K}$. Above this region, the transition zone spans up to the base of the corona ($y = h_\mathrm{c}=15\,\mathrm{Mm}$), where the temperature increases linearly to $T_\mathrm{c}=10^6\,\mathrm{K}$, representing the constant temperature of the corona. 

We consider the atmosphere in hydrostatic equilibrium and current free. The gas pressure distribution is:
\begin{equation}\label{pres}
    {\textstyle  p(y)=\!}
    \left\{
    \begin{array}{rl}
    \begin{alignedat}{2}
      &{\textstyle \!\!p_\mathrm{ch}\exp{\! \left[\frac{\alpha}{T_\mathrm{ch}}\left(\frac{1}{h_\mathrm{ch}+R_{\sun}}-\frac{1}{y+R_{\sun}}\right)\right]}} && 0\leq y < h_\mathrm{ch} \\
     &{\textstyle\! \!p_\mathrm{ch}\exp{\!\left[-\int_{h_\mathrm{ch}}^{y}\frac{\alpha}{T{\scriptstyle(y')}}(R_{\sun}+y')^{-2} dy'\right]}} \quad &&  h_\mathrm{ch}\leq y < h_\mathrm{c}\\
     &{\textstyle \! \!\frac{k_B}{N_Am_i}T_\mathrm{c}n_\mathrm{c}\exp{\!\left[-\frac{\alpha}{T_\mathrm{c}}\left(\frac{1}{h_\mathrm{c}+R_{\sun}}-\frac{1}{y+R_{\sun}}\right)\right]}}  &&  h_\mathrm{c} \leq y ,

    \end{alignedat}
  \end{array} \right.  
\end{equation}
where $\vec{g} =\frac{-G M_\sun}{(y + R_\sun)^2}\vec{e}_y$, $G$ is the gravitational constant, $M_\sun$ is the solar mass, $R_\sun$ is the solar radius, and,
\begin{equation*}
   {\textstyle p_\mathrm{ch}(y)=\frac{k_B}{N_Am_i}T_\mathrm{c}n_\mathrm{c}\exp{\left[\int_{h_\mathrm{ch}}^{h_\mathrm{c}}\frac{\alpha}{T(y')}(R_{\sun}+y')^{-2} dy'\right]}} \, ,
\end{equation*}
where $n_\mathrm{c}=3\times10^8\,\mathrm{cm}^{-3}$ is the number density at height $y=h_\mathrm{c}$, $\alpha= \frac{m_i G M_\sun}{2k_B}$, and $N_A$ is the Avogadro number. With these values, we guarantee that $\beta<1$ in the majority of the volume for all simulations, except at the HS apex where the field strength is zero.   
   
Initially, the temperature within the MFR is equal to that of the corona. Its internal pressure is obtained by proposing a solution close to the equilibrium:
\begin{align} \label{e:presion}
   p_\text{\tiny{FR}}(x,y) = p(y)& +\tfrac{1}{c}\int_{R}^{r+\frac{\Delta}{2}}B_\phi{\scriptstyle(R')}j_z{\scriptstyle(R')}dR'\nonumber\\
   &-\tfrac{1}{c}\int_{R}^{r+\frac{\Delta}{2}}B_{\text{z}}{\scriptstyle(R')}j_\phi{\scriptstyle(R')}dR'.
\end{align}
The associated plasma densities are obtained from the adopted equation of state, i.e.:
\begin{equation}
  {\textstyle \rho=\frac{m_i p(y)}{2k_BT(y)}}.    
\end{equation}

\subsection{Initial configurations}
\label{sec:initial}
\begin{figure*}[h!]
   \centering
   \includegraphics[width=0.8\textwidth]{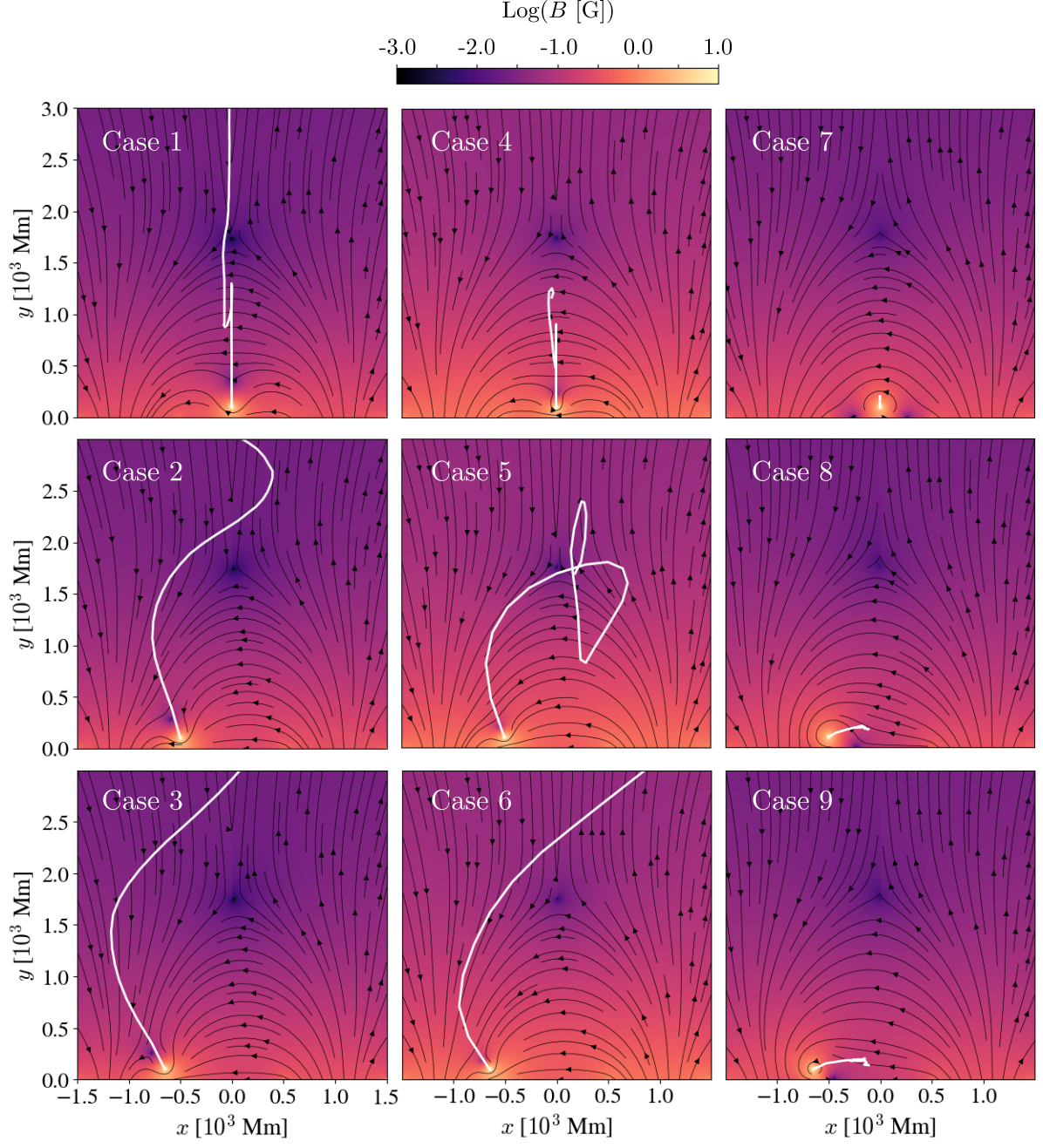}
      \caption{Magnetic configuration at the initial times for the cases. In colour, we plot the magnitude of the magnetic field. The projection of the magnetic field lines into the $xy$-plane are represented by black lines and} the trajectories of the MFR are represented by white lines.
         \label{fig.initial}
\end{figure*}

To analyse the evolution of the MFR, we create several scenarios (see Table~\ref{tab:cases}) in which its position ($x_{\mathrm{MFR}}$) is changed along the HS base. Additionally, we modify the environmental magnetic strength through the parameter $B_0$. The HS field contains a null point of magnetic energy at the interface between open and closed fluxes, specifically at the base of the current sheet. The interplay between the MFR and HS fields induces the emergence of a secondary null point, arising from the mutual flux cancellation of these structures. The location of the secondary null point varies based on the position of the MFR, consequently altering the dynamic behaviour of the eruption, as elaborated in the next section. This can be seen in Fig.~\ref{fig.initial}, which shows the initial magnetic configuration (black lines) and the trajectories of the MFR (white lines). In this figure, dark colours represent the regions of low magnetic energy. 

\begin{figure}[h!]
   \centering
   \includegraphics[width=0.35\textwidth]{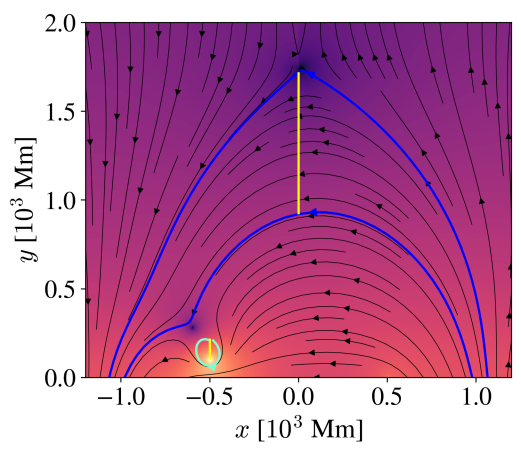}
      \caption{Magnetic flux calculation. The black lines represent the projection of the magnetic field lines into the $xy$-plane magnetic field lines. The blue lines delineate the MC, while the closed light green line symbolises the MFR. To ascertain the magnetic flux, integration is performed along the vertical yellow lines.}
         \label{fig.fluxcalcu}
\end{figure}

Also, considering the results obtained by \cite{2022A&A...662A.113S}, we calculated the magnetic flux of the MFR and its strapping field. Those closed field lines of the HS that overlie the MFR constitute the strapping field (blue lines displayed in case 5 in Fig.~\ref{fig.fluxcalcu}). These lines form a magnetic cage (MC) that constrain the MFR (represented by a closed light green line in Fig.~\ref{fig.fluxcalcu}) eruption. 
Given the symmetry axis, we calculated the poloidal magnetic flux per unit length, for both the MFR and the MC, using:
\begin{equation}
    \phi_{B_p} =\tfrac{1}{L_z} \int_S |\vec{B_p}\cdot\vec{dA}| = \int_\gamma |B_\perp|\, ds.
\end{equation}
For simplicity we define $\gamma$ over a line of $x=c$, where $c$ is a constant. Consequently, $B_\perp = B_x$ and $ds=dy$.
For the calculation of the MFR flux ($\phi_{B_p,\mathrm{MFR}}$), $c$ is determined by the $x-$position of its centre and $y$ limits of $\gamma$ are defined by the height of the MFR centre and the last closed field line belonging to the MFR (vertical yellow line centred in the MFR shown in Fig.~\ref{fig.fluxcalcu}). To calculate the MC flux, $\phi_{B_p,\mathrm{MC}}$, we chose $c$ and $y$ limits of $\gamma$ are defined by the height of the first and last closed magnetic field lines, tied to the chromosphere, located above the MFR (in Fig.~\ref{fig.fluxcalcu}, the vertical yellow line located at $x=0$).

%-----------------------------------------------------------------

\section{Results}
\label{s:results}

\subsection{Summary of topology and dynamic behaviour}

Below we group the cases according to their reconnection dynamics. We define successful ascent cases as those in which the MFR rises beyond the top of the HS and maintains an upward trajectory until the end of the simulation.
The unsuccessful ascent cases can be further split into \emph{failed ascents} where the ascent starts but then stalls and \emph{confined ascents} where the ascent doesn't begin to start with. 

\begin{figure*}[]
   \centering
   \includegraphics[width=0.9\textwidth]{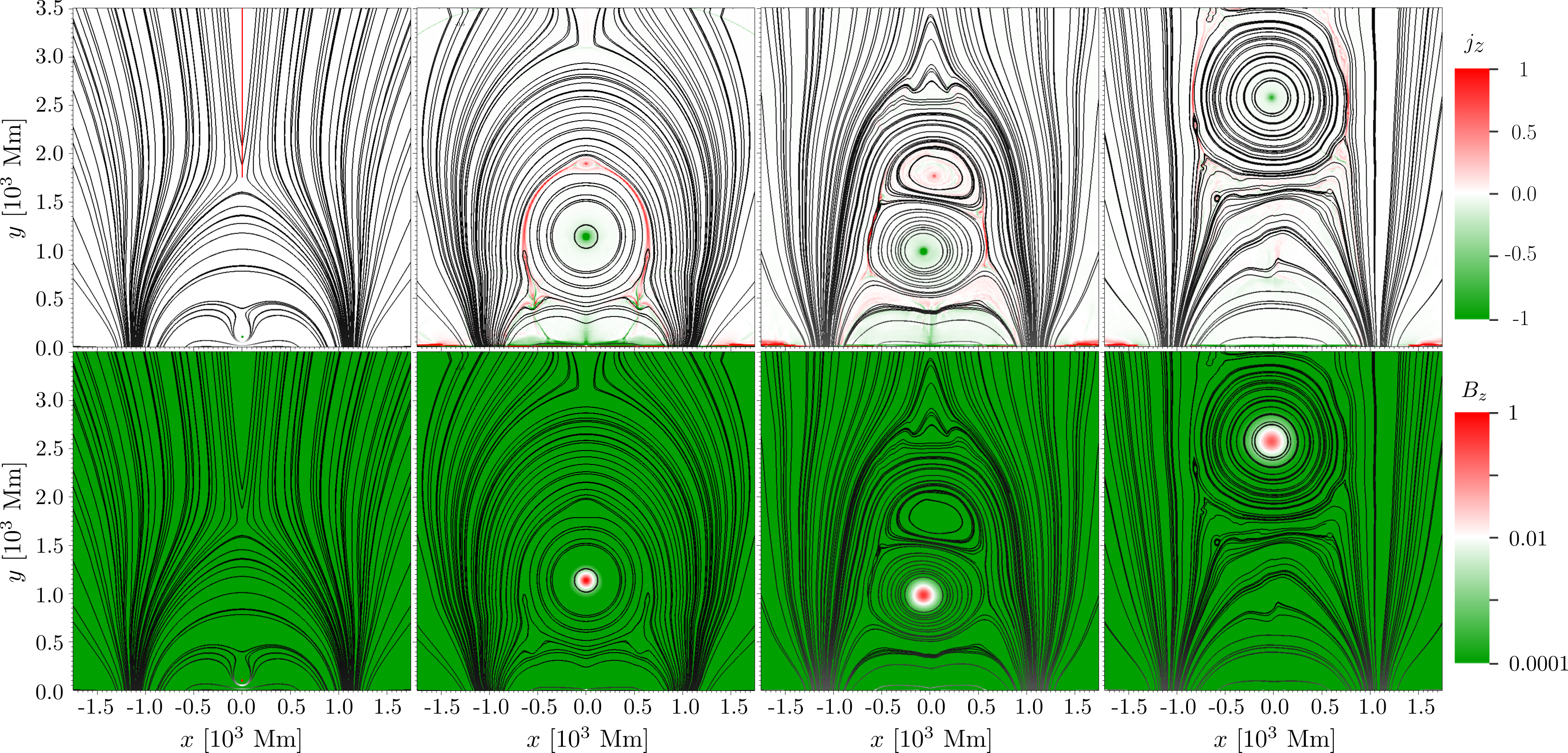}
      \caption{
      Upper panel: Evolution of the current $j_z$ (in units of $10^{-10}\,\textrm{G\,cm}^{-1}$) for case 1. Bottom panel: Evolution of $B_z$ (in units of $\mathrm{G}$) within the MFR. The initial time is displayed in the first panel, $t=5000~\mathrm{s}$ in second panel, $t=16000~\mathrm{s}$ in third panel and $t=17000~\mathrm{s}$ in last panel for both upper and bottom panels for case 1. The black lines represent the projection of the magnetic field lines into the $xy$-plane. 
      }
         \label{fig.eruption2}
\end{figure*}

%\noindent
\subsubsection*{Cases 1 \& 4} 
In these cases, the MFR is centred inside the HS ($x_{\mathrm{MFR}}=0$) and the MFR polarity is inverted compared to the HS polarity. In both, the combination of MFR and HS magnetic fields creates a secondary null point above the MFR (see Fig \ref{fig.initial}). However, the background HS field strength differs by a factor of two (Table \ref{tab:cases}) between the cases. 

\begin{figure}
    \centering
    \includegraphics[width=0.36\textwidth]{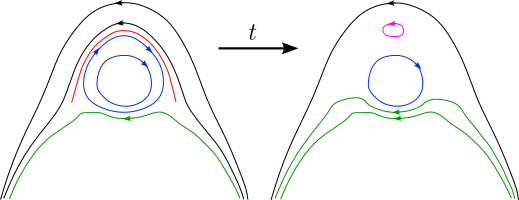}
    \caption{Diagram illustrating the magnetic reconnection occurring in cases 1 and 4. Black lines represent the magnetic field lines of the MC, blue lines indicate the magnetic field lines of the MFR, green lines depict the arcade below the MFR, and the pink line represent the magnetic island formed. The red line highlights the current sheet region between the MFR and the MC.}
    \label{fig:scheme1}
\end{figure}

Case 1 with the weaker HS field is a successful ascent (see the evolution in the animated version). Figure~\ref{fig.eruption2} displays the evolution of the current $j_z$ and the $z$-component of the magnetic field. Initially, as the MFR rises, the null point collapses into a current sheet that becomes stretched around the flux rope, see the second column of the figure. The change in connectivity as a result of reconnection within this current layer is shown schematically in Fig.~\ref{fig:scheme1}.
Reconnection in this current sheet acts to reduce the overlying strapping field (the outer magnetic field of MFR (blue line) in left panel of the figure, reconnects with the overlying magnetic field of MC (black line) which converts into a new magnetic field line below the MFR (new green line in right panel). In effect, the reconnection ahead of the MFR allows it to ``tunnel'' through the strapping field at the expense of loosing outer layers of poloidal magnetic flux in the process. 

This reconnection is similar to that of the magnetic breakout model \citep{Antiochos1999} for CME eruptions. However, the reconnection in our model differs from classic breakout reconnection in some key ways. Firstly, in the breakout model the initial null point is a feature of a quadrapolar surface magnetic flux distribution whereas here we have a background flux distribution which is bipolar. Secondly, in the breakout model the presence of the four flux regions in the background magnetic field means that strapping field is reconnected into two adjacent ``lobe'' flux systems. This facilitates the eruption purely by reducing the overall downward magnetic tension on the flux rope/filament channel. Here the strapping field is moved behind the flux rope, reducing the magnetic tension in the same manner but also adding additional magnetic pressure/tension behind the flux rope which further drives it upwards. However, in the breakout model a flare current layer also forms behind the flux rope which our model lacks. Reconnection in the flare current layer increases the magnetic flux of the flux rope, aiding in the eruption. Nevertheless, despite the differences there are still clear similarities between the evolution we see in our simulation and previous breakout simulations which will be discussed further below. We also note that \cite{Titov2022} identified similar breakout-like reconnection in a bipolar magnetic configuration, referring to it as ``breakthrough'' reconnection. However, their configuration also contained a flare current layer so is again not quite the same as the reconnection dynamics in these simulations. For simplicity we will refer to the reconnection in our simulations as \emph{null point} reconnection, but will highlight the similarities to previous breakout simulations where appropriate.

Returning to the evolution in case 1, not long after the onset of reconnection in the current sheet ahead of the MFR, a magnetic island forms which quickly grows to become comparable in size to the MFR itself (Fig.~\ref{fig.eruption2} central panels). The field configuration is also shown schematically in Fig.~\ref{fig:scheme1}. The large size that the island grows to is an artefact of the symmetry of our system, see e.g. \citet{MacNeice2004}. The formation of the island inhibits the initial rise of the flux rope, which then experiences a bounce-back before eventually ascending (Fig.~\ref{fig.eruption2}, right panels). 

\begin{figure*}[]
   \centering
   \includegraphics[width=0.9\textwidth]
   {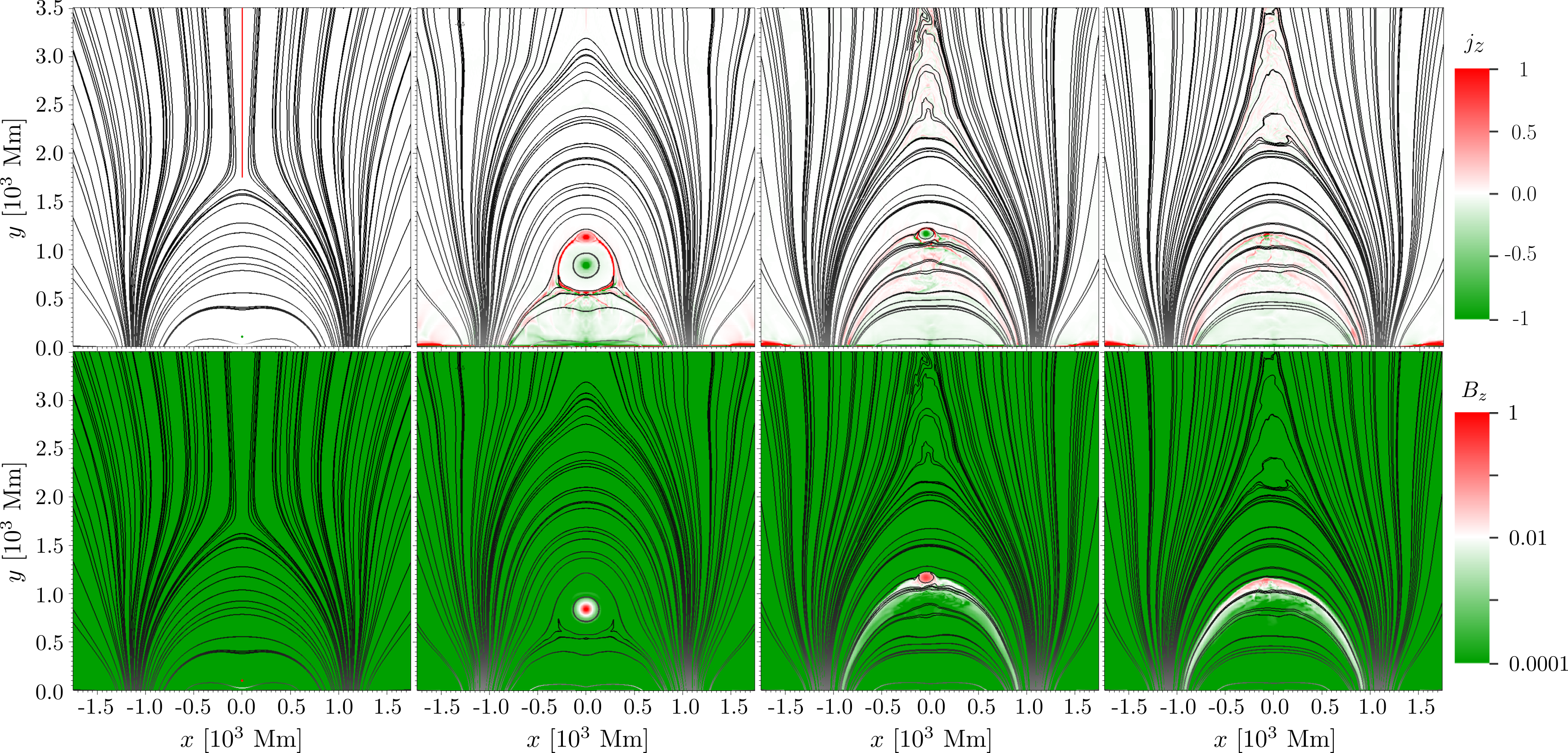}
      \caption{Same as Fig. \ref{fig.eruption2} for case 4.  
      The initial time is displayed in the first panel, $t=5000~\mathrm{s}$ in second panel, $t=16000~\mathrm{s}$ in third panel and $t=17000~\mathrm{s}$ in last panel for both upper and bottom panels. }
         \label{fig.nonerupt1}
\end{figure*}

Case 4 has a similar initial evolution, but with much less initial expansion of the MFR due to the stronger HS field. In this case the ascent fails when the stronger strapping field absorbs the initial rise of the MFR, which sets up an oscillation of the MFR beneath the apex of the HS (see the trajectory in Fig. \ref{fig.initial}). Also, the null point reconnection erodes not just the outer flux surfaces of the MFR, but continues all the way to its core. Eventually, the MFR is destroyed entirely (see evolution of $j_z$ in upper panels of Fig.~\ref{fig.nonerupt1}, and in the animated version). A by-product of this is that the axial flux of the flux rope becomes spread out along closed loops within the HS (bottom panels of Fig.~\ref{fig.nonerupt1}). Again, there are parallels with previous work on jets and confined eruptions invoking breakout reconnection where in those cases the breakout reconnection proceeds until it reaches the core of the erupting flux rope/filament channel, transferring its twist to adjacent flux systems \citep[e.g.][]{Devore2008,Doyle2019}.

%\noindent
\subsubsection*{Cases 2, 3, 5, and 6}
In all these cases the MFR's initial position is shifted to the left. Again the MFR has a magnetic polarity that is opposite to that of the HS but now the null point is initially present above and to the left of the MFR (Fig.~\ref{fig.initial}). The MFR in cases 3 and 6 is placed further to the left than in cases 2 and 5, and within each pairing two HS field strengths are considered, see Table~\ref{tab:cases} for details. 

\begin{figure}
    \centering
    \includegraphics[width=0.37\textwidth]{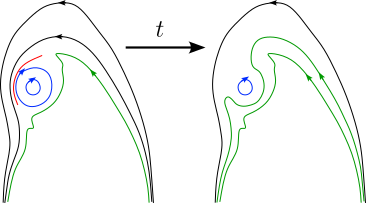}
    \caption{Diagram illustrating later magnetic reconnection occurring in cases 2, 3, 5 and 6. Black lines represent the magnetic field lines of the MC, blue lines indicate the magnetic field lines of the MFR, and green lines depict the arcade below the MFR. The red line highlights the current sheet region between the MFR and the MC.}
    \label{fig:scheme3}
\end{figure}

\begin{figure*}[]
   \centering
   \includegraphics[width=0.9\textwidth]{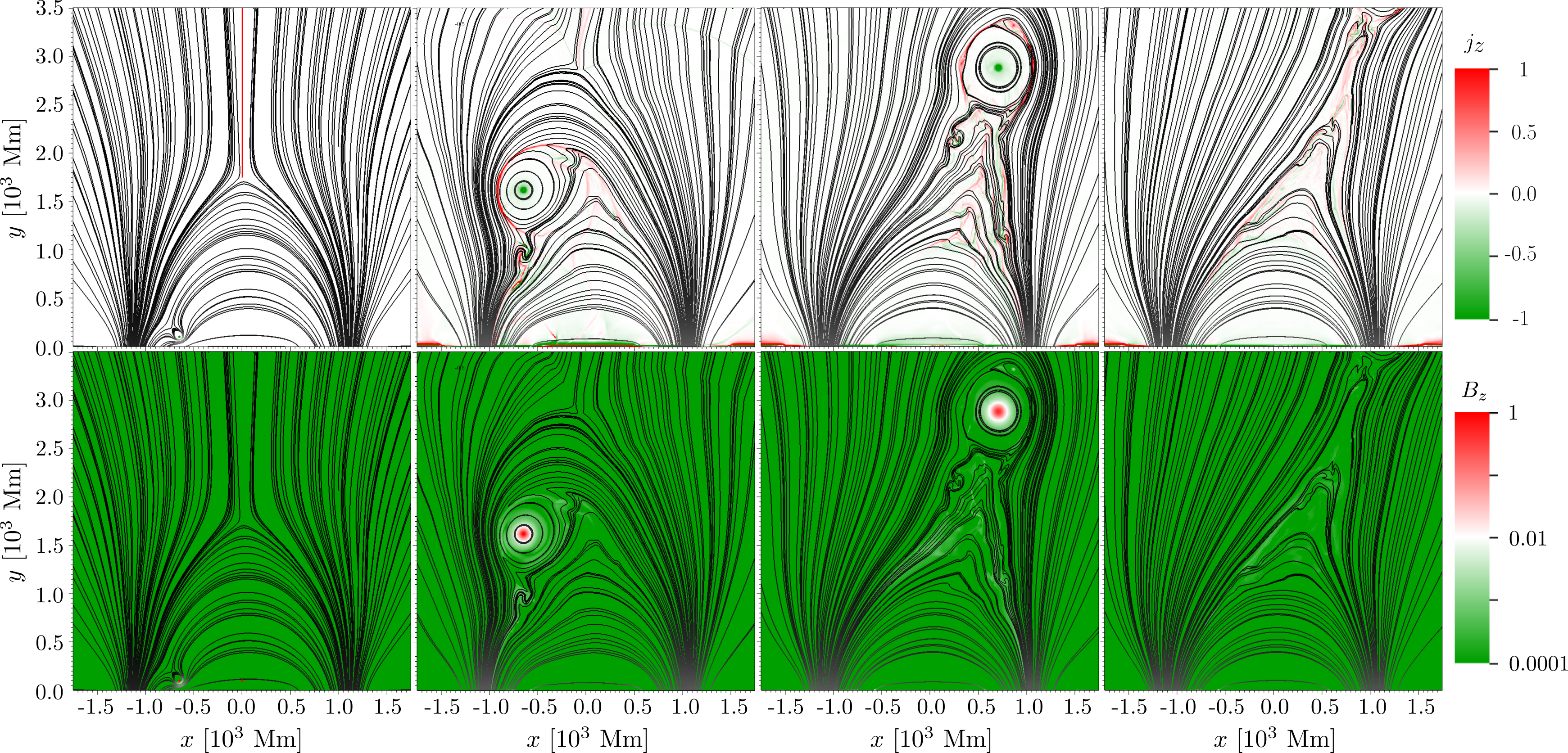}
      \caption{Same as Fig. \ref{fig.eruption2} for case 6. The initial time is displayed in the first panel, $t=5000~\mathrm{s}$ in second panel, $t=10000~\mathrm{s}$ in third panel and $t=15000~\mathrm{s}$ in last panel for both upper and bottom panels.
      }
         \label{fig.eruption}
\end{figure*}

In all cases the MFR initially rises towards the null point, deflecting the eruption towards the left flank of the HS before being redirected back towards the apex of the HS (see the trajectories in Fig.~\ref{fig.initial}). As in the previous cases, the expansion of the MFR collapses the null point into a current sheet, inducing reconnection. The flux transfer induced by this reconnection is shown schematically in Fig.~\ref{fig:scheme3}. Due to the asymmetry of the system, when islands form in the current layer they are quickly ejected so do not grow to sizes large enough to affect the overall ascent. Cases 2, 3 and 6 are successful ascent and follow a similar evolution (see the animated versions). Figure~\ref{fig.eruption} shows the evolution of case 6 as an example (see also the animated version). The induced reconnection tunnels the MFR towards the open flux, discussed further in \S \ref{sec:flux}, reducing the overall strapping flux and ultimately allowing the MFR to continue on its upward trajectory when it reaches the apex of the HS (Fig.~\ref{fig.eruption}, right panels). 

\begin{figure*}[h!]
   \centering
   \includegraphics[width=0.9\textwidth]{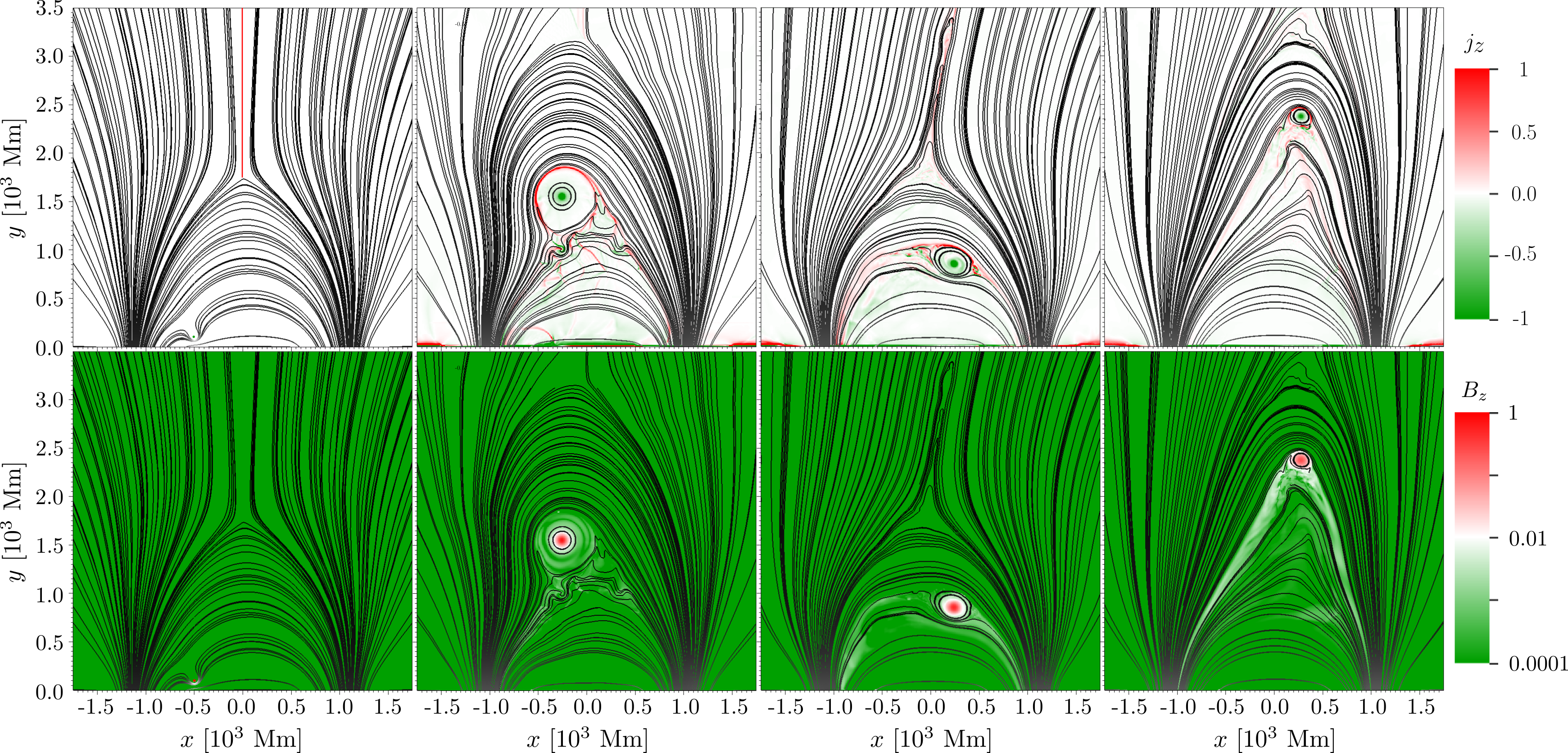}
      \caption{
      Same as Fig. \ref{fig.eruption2} for case 5. The initial time is displayed in the first panel, $t=5000~\mathrm{s}$ in second panel, $t=15000~\mathrm{s}$ in third panel and $t=24000~\mathrm{s}$ in last panel for both upper and bottom panels.}
         \label{fig.nonerupt3}
\end{figure*}

By contrast, the ascent in case 5 fails. In this case, as in case 4, the stronger overlying field absorbs the initial rise of the MFR, redirecting it downwards where it then bounces back and forth within the HS. Eventually, the null point reconnection reaches the core of the MFR, spreading its axial field component along the HS field, see Fig.~\ref{fig.nonerupt3} and the animated version.

\subsubsection*{Cases 7, 8 and 9}
\begin{figure*}[h!]
   \centering
   \includegraphics[width=0.9\textwidth]{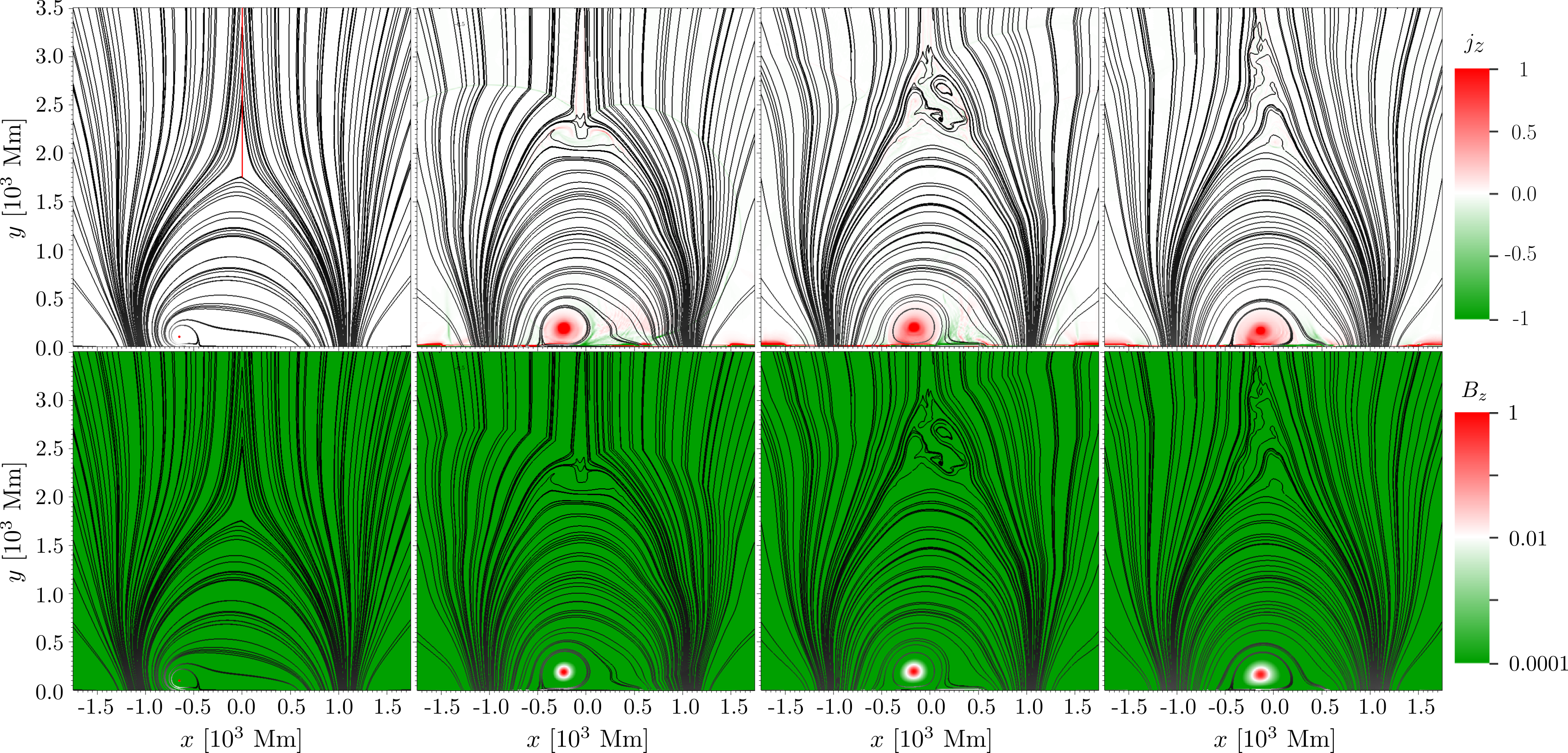}
      \caption{
      Same as Fig. \ref{fig.eruption2} for case 9. The initial time is displayed in the first panel, $t=5000~\mathrm{s}$ in second panel, $t=10000~\mathrm{s}$ in third panel and $t=25000~\mathrm{s}$ in last panel for both upper and bottom panels.}
         \label{fig.nonerupt2}
\end{figure*}

In these cases the MFR has the same polarity as the HS magnetic field. In case 7, where the MFR is positioned directly below the apex of the HS, the symmetry of the system leads to two null points either side of the MFR on the lower boundary. In cases 8 and 9 where the MFR is displaced to the left, a single null point initially resides below and to the right of it (Fig.~\ref{fig.initial}). No ascent occurs in these cases, and the MFRs remain confined to near their initial position (see animated versions). The evolution in case 9 is shown in Fig.~\ref{fig.nonerupt2} as an example. The MFR first expands and moves to the right, before relaxing to a new equilibrium. 

\subsection{Analysis of magnetic fluxes vs ascent behaviour}
\label{sec:flux}
In this section we analyse the losses and ratios of the magnetic fluxes per unit length of the MFR and the MC. This analysis will help quantify the ascent behaviour of the different systems. 

\begin{table*}[h!]
     $$ 
         \begin{array}{lccccc}
            \hline
            \noalign{\smallskip}
            \mathrm{Case }     &  x_{\mathrm{MFR}}\ &  B_0  &  \phi_{B_p,\mathrm{MC}}  & \phi_{B_p,\mathrm{MFR}}  & \mathrm{Erup.}   \\
            \noalign{\smallskip}
            \hline

\hline
             1 &  0.0  & 0.1 & 58    & -133 &   y \\
             2 & -0.5  & 0.1 & 27    & -128 &    y   \\
             3 & -0.65 & 0.1 & 2     & -130 &   y  \\
\hline
             4 &  0.0  & 0.2 & 165  & -102 &   n   \\
             5 & -0.5  & 0.2 & 112  & -100 &   n     \\
             6 & -0.65 & 0.2 & 59    & -100 &   y   \\
\hline
             7 &  0.0  & 0.1 & 158    & 193 &   n \\
             8 & -0.5  & 0.1 & 155    & 184 &   n   \\
             9 & -0.65 & 0.1 &  151  & 163 &   n  \\
            \noalign{\smallskip}
            \hline
         \end{array}
     $$ 
      \caption[]{Magnetic fluxes of the different cases. The units for MFR positions $x_{\mathrm{MFR}}$ are expressed in $10^3\,\mathrm{Mm}$, magnetic fields $B_0$ in $\mathrm{G}$, poloidal magnetic fluxes per unit length $\phi_{B_p}$ in $10^8\,\mathrm{Mx\,cm^{-1}}$. 
      The last column indicates whether the case erupts ($y$) or not ($n$). 
      }
         \label{tab:cases}
   \end{table*}

\begin{figure}[h!]
   \centering
   \includegraphics[width=0.5\textwidth]{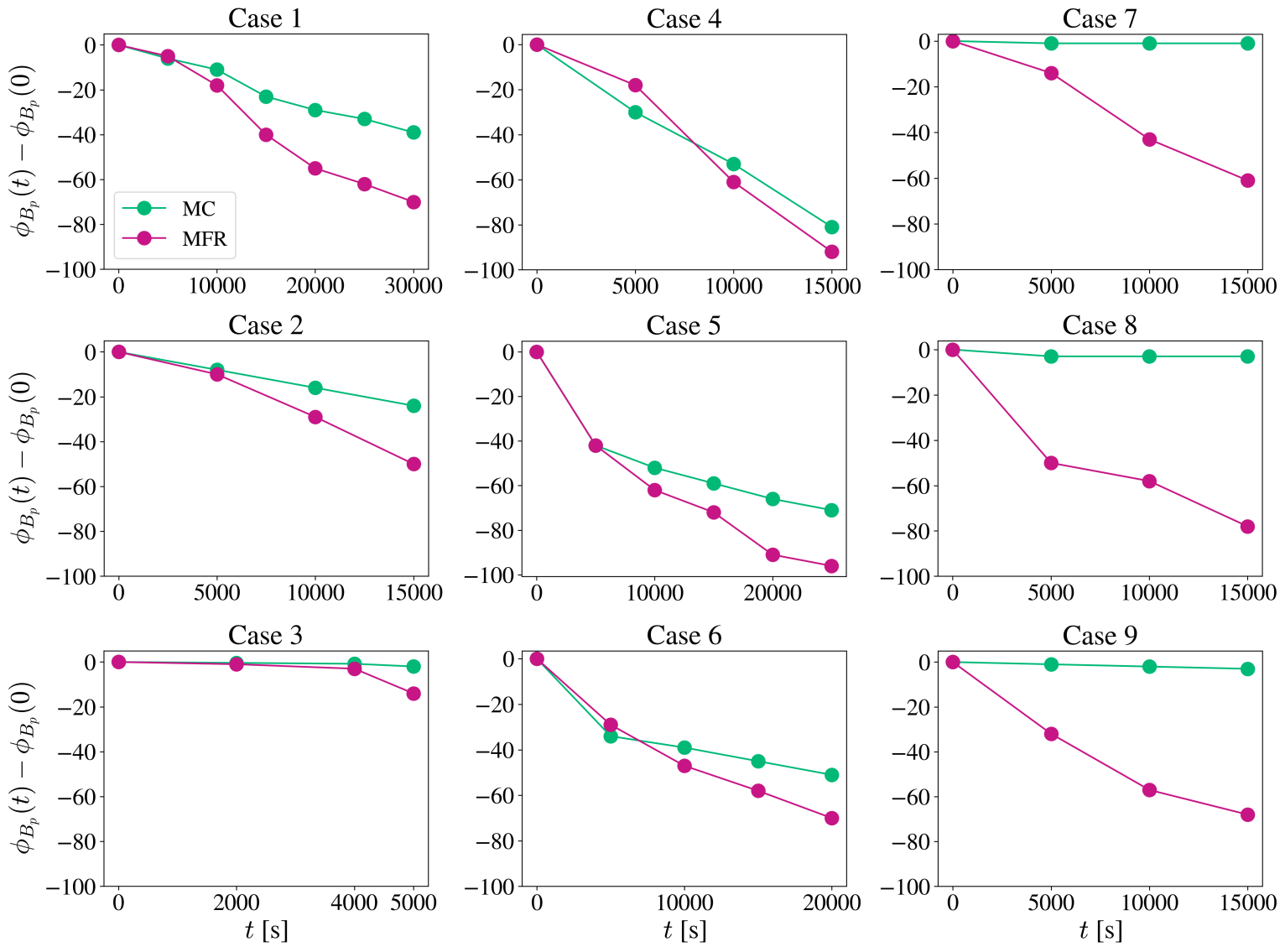}
      \caption{Evolution of the magnetic flux per unit length difference relative to its initial value for both the MC (green lines) and the MFR (pink lines).}
         \label{fig.diffflux}
   \end{figure}

In Table~\ref{tab:cases}, we also show the poloidal (in-plane) magnetic fluxes per unit length ($\phi_{B_p}$) calculated for the MC and MFR at $t=0$ (calculated as described in \S \ref{sec:initial}, see also Fig. \ref{fig.fluxcalcu}). The change in time ($\phi_{B_p}(t)-\phi_{B_p}(0)$) throughout each simulation of both quantities is shown in Fig.~\ref{fig.diffflux}.
It can be seen that both the MC and MFR lose poloidal magnetic flux per unit length over time, with the losses being lower in the MC compared to the MFR. This is due to the competing effects of the reconnection occurring in the system. For cases where the null point is above the MFR initially, the null point reconnection induced by the motion of the MFR reduces the flux of the MFR and MC equally. The reconnected MC flux is transferred behind the MFR as it tunnels upward. However, at the same time reconnection in the HS current sheet adds  some flux to the MC. For cases 1–6, we show a significant decrease in the flux of the MC, indicating that despite this contribution, the null point reconnection plays a predominant role in reducing the flux in the MC. For cases where the null point is below the flux rope axis, the reconnection tunnels the flux rope laterally, and deeper into the closed field, adding flux to the MC from the MFR in addition to the flux added by the HS current sheet reconnection.

\begin{figure}[h!]
   \centering
   \includegraphics[width=0.5\textwidth]{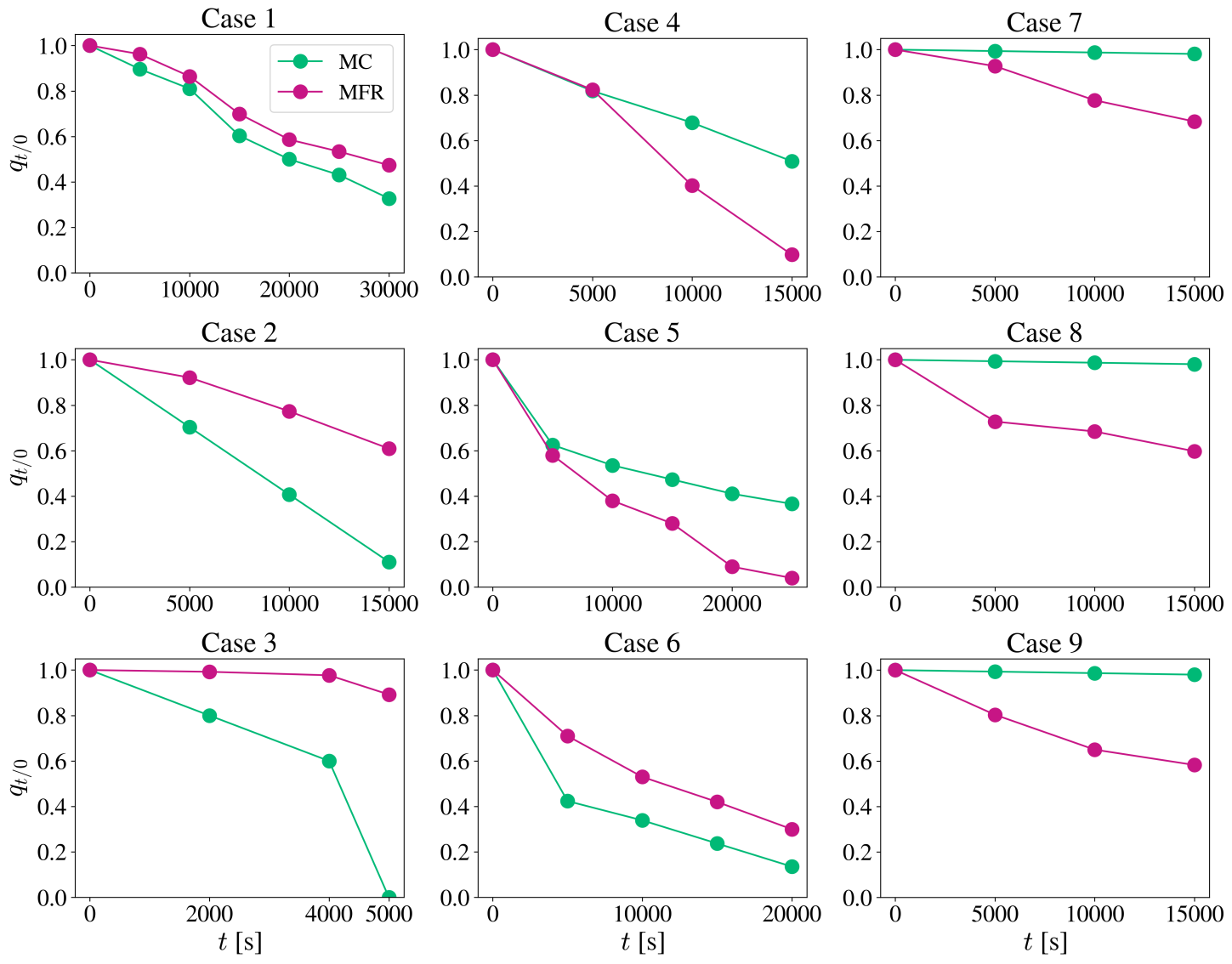}
      \caption{Evolution of magnetic flux per unit length decay rate for the MC (green lines) and MFR (pink lines) over time.}
         \label{fig.ratiosflux}
   \end{figure}

To analyse the evolution of magnetic flux per unit length in both the MC and MFR over time, Fig.~\ref{fig.ratiosflux} shows the absolute values of the ratios of the poloidal magnetic fluxes per unit length between a time ``$t$'' with respect to an initial time ``$0$'' for the MC ($q^{\mathrm{MC}}_{t/0}=\phi_{B_p,\mathrm{MC}}(t)/\phi_{B_p,\mathrm{MC}}(0)$) and for the MFR ($q^{\mathrm{MFR}}_{t/0}=\phi_{B_p,\mathrm{MFR}}(t)/\phi_{B_p,\mathrm{MFR}}(0)$). From here, we observe that the trend of flux decay continues over time, both for the MC ($q^{\mathrm{MC}}_{t/0} <1$) and the MFR ($q^{\mathrm{MFR}}_{t/0} < 1$) in all cases. However, we note that for successful and unsuccessful ascent cases, these decays differ. Unsuccessful ascent cases (cases 4, 5, 7, 8 and 9) show that the decay of the MC flux is smaller than that of the MFR. 
The opposite is observed for successful ascent cases. 

The key difference between the successful and unsuccessful ascent (failed and confined) cases that helps explain this is the initial poloidal flux within the MFR, see Table \ref{tab:cases}. In the successful ascent cases the poloidal flux of the flux rope is a relatively high fraction of the MC flux, leading to a larger relative drop in MC flux than MFR flux as it tunnels outwards. 

\begin{figure}[h!]
   \centering
   \includegraphics[width=0.45\textwidth]{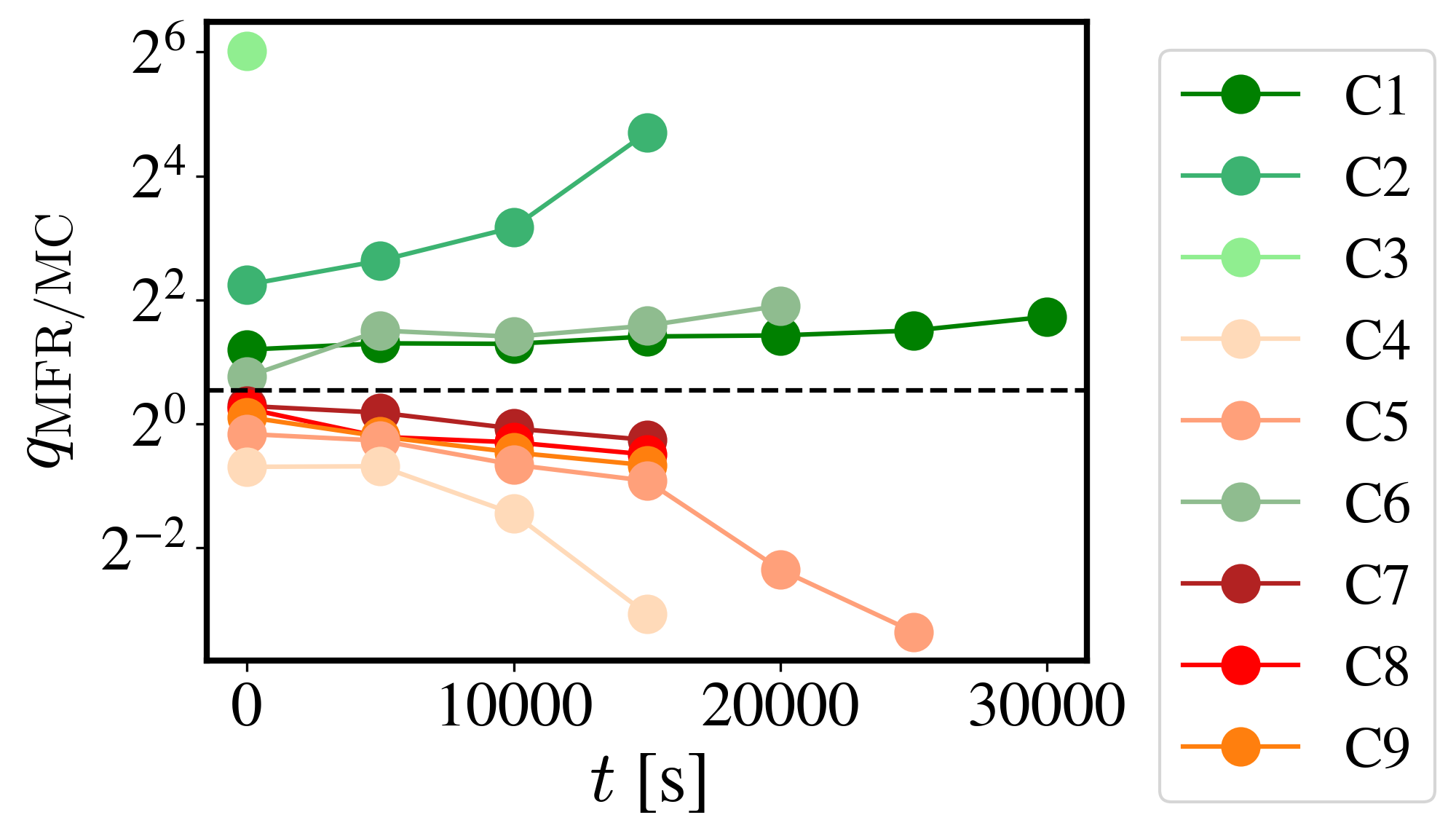}
      \caption{Evolution of the ratio between the MFR and MC magnetic flux per unit length over time. The green colours represent successful ascent cases, while red colours represent unsuccessful ascent cases. The black dashed line marks the threshold of 1.5.}
         \label{fig.ratios}
   \end{figure}

This can be seen more clearly when the fluxes of the MFR and MC are compared with time. Fig.~\ref{fig.ratios} shows the ratio between the magnetic flux per unit length of the MFR and the MC ($q_{\mathrm{MFR}/\mathrm{MC}}=\phi_{B_p,\mathrm{MFR}}(t)/\phi_{B_p,\mathrm{MC}}(t)$) over time. Once again, there are distinct differences in behaviour between unsuccessful and successful ascent cases. In unsuccessful ascent cases, the ratios decrease with time indicating a relative increase in strapping/MC flux, while the successful ascent cases exhibit an ascending trend.
This is indicative of a relative drop in the strapping/MC flux relative to the flux of the flux rope. This is also a typical feature of breakout-type eruptions \citep[e.g.][]{Antiochos1999} and indicates that in a similar manner the null point reconnection in our simulations is playing a key role in facilitating these successful ascent through the reduction of the overlying flux.
The transition from unsuccessful to successful ascent occurs around $q_{\mathrm{MFR}/\mathrm{MC}} \sim 1.5$. With the successful ascent cases having $q_{\mathrm{MFR}/\mathrm{MC}} \gtrsim 1.5$ throughout their evolution. 
\\

\noindent

%-----------------------------------------------------------------

\section{Conclusions}
\label{s:conclus}

In this work we have considered the role of the HS topology on the ascent behaviour and deflection of MFR eruptions. The limited parameter study performed was broad enough to include both stable and unstable cases and with differing levels of asymmetry in the flux rope location. 

To analyse the ascent behaviour, as mentioned in \S \ref{sec:magfield}, we selected an non-equilibrium configuration for the MFR immersed in an HS background. This setup results in an early deceleration of the MFR, with the interaction of the flux rope with its surroundings being the focus of our study. Although this model cannot fully reproduce the initial onset of the ejection (slow-to-fast rise phase), we define a successful ascent as the MFR reaching beyond $2.5 R_{\odot}$, which remains a reasonable and consistent threshold within the framework of our analysis. 

Regarding the flux rope deflection, our findings confirm previous results in showing that the MFR is deflected towards regions of weaker magnetic field guided by the topology of the magnetic field lines \citep[e.g.,][]{2023AdSpR..72.5263W,2023ApJ...957...74B,2023ApJ...953..150S}.
In the HS topology this region corresponds to the heliospheric current sheet. In cases where a null point was present above the flux rope initially, the null point also served as a low magnetic field region to deflect towards. As we shown in, for example, case 3, the combination of both features can lead to a change in deflection direction resulting in a curved path for the flux rope during the eruption. 
A comparable behaviour was noted in \cite{2018ApJ...862...86Y}. A filament positioned adjacent to a heliospheric streamer initially underwent a northward eruption and subsequently deflected southward towards the magnetic minimum in the streamer configuration. However, we note that our simulations are 2.5D and this pre-supposes that 3D effects such as twist transfer along field lines does not occur which can also lead to deflections \citep{Wyper2021}.

Null point reconnection is shown to play both a stabilising and destabilising role in the ascent, depending on the configuration. Reconnection occurred in our simulations when the null point was initially above the inserted flux rope which then rapidly collapsed into a current layer once the simulation was underway. The reconnection in this layer acts to erode the outer poloidal flux of the flux rope while also reducing the strapping flux of the magnetic cage. When the strapping flux of the magnetic cage is sufficiently great, the initial ascent of the MFR stalls and the MFR then oscillates in position while the reconnection erodes the flux of the flux rope until it is destroyed (cases 4 \& 5).

However, in other instances (cases 1, 2, 3 \& 6) the reconnection acted to enable the ascent. In such cases the poloidal flux of the flux rope was initially large enough that the erosion of poloidal flux did not reach the core of the flux rope, destroying it as in the cases with failed ascents. Instead, the flux rope core maintains its coherence as the null point reconnection tunnelled it through the strapping flux. In the asymmetric cases (2, 3 \& 6) this brought it closer to the open-closed boundary and ultimately into the open field in a manner similar to that shown by \citet{2013ApJ...771...82M} for a breakout-type eruption. While when the configuration was symmetric, the reconnection simply reduced the overlying flux enough that the flux rope became unstable (case 1). Finally, in the absence of reconnection when the null point is initially beneath or adjacent to the inserted flux rope (cases 7, 8 \& 9) we find that the flux rope motion is rapidly stabilised.

At this point, it is appropriate to mention that the symmetry of the system in certain simulations, such as case 1, can lead to the generation of an artificially large island ahead of the flux rope. These islands likely hinder the ascent by suppressing the null-point/breakout reconnection. However, the fact that the flux rope still rises despite this suppression suggests that it is particularly unstable. We also note that such large islands have been seen in previous studies \citep[e.g.][]{MacNeice2004}.

As has been noted in previous work \citep[e.g.][]{Amari2018,2022A&A...662A.113S}, we also found that the magnetic flux of the MC plays a crucial role in the ascent process. This occurs both through the bodily confinement of the flux rope and also through dictating the extent of the reconnection dynamics. Our analysis of the magnetic fluxes within the system and their evolution in time demonstrated a strong correlation between the ascent occurrence and the ratio of the magnetic fluxes in the MFR and MC. Although it does not directly predict the eruptive nature of the MFR, this ratio provides valuable insights that contribute to a deeper understanding of its evolution and the conditions influencing its behaviour. We identified a critical value in this ratio that serves as a distinguishing factor for ascent outcomes. Specifically, when the poloidal flux of the MFR is greater than $1.5$ times the strapping flux of the MC ($q_{\textrm{MFR/MC}}>1.5$), an ascent is triggered. However, we note that our simulations lacked a solar wind, which would be expected to open high-lying magnetic field lines, aiding the ascent.
Thus, the threshold we have identified in $q_{\textrm{MFR/MC}}$ for HSs is likely an upper bound.

Another consequence of the absence of solar wind emulation is the unbalanced reconnection in the HCS. As mentioned in \S \ref{sec:flux}, magnetic field lines reconnect at the apex of the HS, contributing to the flux of the strapping field. Were a solar wind included, this addition of magnetic flux would be balanced on average by the re-opening of closed field pulled out by the solar wind \citep[e.g.][]{Higginson2017}.
Nonetheless, our results indicate that the flux contribution from this reconnection to the magnetic cage is negligible over the time frame of the eruptions we investigated. In all cases, the null point reconnection plays a dominant role, and the strapping flux consistently decreases.

It is important to note that the threshold for flux ratios depends implicitly on the position of the null point. The time evolution captures this dynamic, as the cases with the null point below/adjacent to the MFR have a ratio which decreases over time. We can conclude that while the null point's position is crucial, it is not sufficient by itself. The flux ratio must also exceed the threshold. For future work, configurations should be considered 
where the null point is below a MFR which is unstable to ascent (which we have not explored here) to see if the flux ratio remains robust in those cases. However, this falls outside the current scope of our study, which focuses on cases dominated by reconnection above/adjacent to the flux rope.

In future, it will also be interesting to consider how this threshold could be extended to three dimensions, or how it relates to other thresholds for ascent behaviour based on the magnetic configuration that have been identified in other studies \citep[e.g.][]{2022FrASS...9.9135R}. Regardless of the specifics, what is clear is that to understand flux rope ascents it is crucial to take into account not just the properties of the flux rope, but also how it interacts with its magnetic surroundings in order to be able to reliably predict whether an eruption will be successful.

\begin{acknowledgements}
     MC and GK are members of the Carrera del Investigador Cient\'ifico (CONICET). MC knowledges support from the DynaSun project, which has received funding under the Horizon Europe programme of the European Union, grant agreement no. 101131534. Views and opinions expressed are however those of the author only and do not necessarily reflect those of the European Union and therefore the European Union cannot be held responsible for them. PFW was supported by STFC (UK) consortium grant ST/W00108X/1 and a Leverhulme Trust project grant. AS was supported by an appointment to the NASA Postdoctoral Program at the NASA Goddard Space Flight Center, administered by Oak Ridge Associated Universities under contract with NASA. MC, GK and AS acknowledge support from CONICET under grant PIP  number 11220200103150CO. OEKR was supported by a UKRI/STFC PhD studentship. We thank Anthony Yeates for useful discussions. MC also acknowledges the hospitality of the Department of Mathematical Sciences were part of this work was carried out.
     Also, we thank the Centro de C\'omputo de Alto Desempe\~no (UNC), where the simulations were performed. The software used in this work was developed in part by the DOE NNSA- and DOE Office of Science-supported Flash Center for Computational Science at the University of Chicago and the University of Rochester. 
 
\end{acknowledgements}

   \bibliographystyle{aa} % style aa.bst
   \bibliography{biblio} % your references Yourfile.bib

\begin{thebibliography}{29}
\expandafter\ifx\csname natexlab\endcsname\relax\def\natexlab#1{#1}\fi

\bibitem[{{Amari} {et~al.}(2018){Amari}, {Canou}, {Aly}, {Delyon}, \&
  {Alauzet}}]{Amari2018}
{Amari}, T., {Canou}, A., {Aly}, J.-J., {Delyon}, F., \& {Alauzet}, F. 2018,
  \nat, 554, 211

\bibitem[{{Antiochos} {et~al.}(1999){Antiochos}, {DeVore}, \&
  {Klimchuk}}]{Antiochos1999}
{Antiochos}, S.~K., {DeVore}, C.~R., \& {Klimchuk}, J.~A. 1999, \apj, 510, 485

\bibitem[{{Ben-Nun} {et~al.}(2023){Ben-Nun}, {T{\"o}r{\"o}k}, {Palmerio},
  {Downs}, {Titov}, {Linton}, {Caplan}, \& {Lionello}}]{2023ApJ...957...74B}
{Ben-Nun}, M., {T{\"o}r{\"o}k}, T., {Palmerio}, E., {et~al.} 2023, \apj, 957,
  74

\bibitem[{{DeVore} \& {Antiochos}(2008)}]{Devore2008}
{DeVore}, C.~R. \& {Antiochos}, S.~K. 2008, \apj, 680, 740

\bibitem[{{Doyle} {et~al.}(2019){Doyle}, {Wyper}, {Scullion}, {McLaughlin},
  {Ramsay}, \& {Doyle}}]{Doyle2019}
{Doyle}, L., {Wyper}, P.~F., {Scullion}, E., {et~al.} 2019, \apj, 887, 246

\bibitem[{{Forbes}(1990)}]{1990JGR....9511919F}
{Forbes}, T.~G. 1990, Journal of Geophysical Research, 95, 11919

\bibitem[{{Forbes}(2000)}]{2000JGR...10523153F}
{Forbes}, T.~G. 2000, \jgr, 105, 23153

\bibitem[{{Fryxell} {et~al.}(2000){Fryxell}, {Olson}, {Ricker}, {Timmes},
  {Zingale}, {Lamb}, {MacNeice}, {Rosner}, {Truran}, \&
  {Tufo}}]{2000ApJS..131..273F}
{Fryxell}, B., {Olson}, K., {Ricker}, P., {et~al.} 2000, The Astrophysical
  Journal Supplement Series, 131, 273

\bibitem[{{Higginson} {et~al.}(2017){Higginson}, {Antiochos}, {DeVore},
  {Wyper}, \& {Zurbuchen}}]{Higginson2017}
{Higginson}, A.~K., {Antiochos}, S.~K., {DeVore}, C.~R., {Wyper}, P.~F., \&
  {Zurbuchen}, T.~H. 2017, \apj, 837, 113

\bibitem[{{Hu}(2001)}]{2001SoPh..200..115H}
{Hu}, Y.~Q. 2001, \solphys, 200, 115

\bibitem[{{Karpen} {et~al.}(2024){Karpen}, {Kumar}, {Wyper}, {DeVore}, \&
  {Antiochos}}]{Karpen2024}
{Karpen}, J., {Kumar}, P., {Wyper}, P., {DeVore}, C.~R., \& {Antiochos}, S.
  2024, \apj in press

\bibitem[{{Krause}(2019)}]{2019A&A...631A..68K}
{Krause}, G. 2019, \aap, 631, A68

\bibitem[{{Krause} {et~al.}(2018){Krause}, {C{\'e}cere}, {Zurbriggen}, {Costa},
  {Francile}, \& {Elaskar}}]{2018MNRAS.474..770K}
{Krause}, G., {C{\'e}cere}, M., {Zurbriggen}, E., {et~al.} 2018, \mnras, 474,
  770

\bibitem[{{Lee} \& {Deane}(2009)}]{2009JCoPh.228..952L}
{Lee}, D. \& {Deane}, A.~E. 2009, Journal of Computational Physics, 228, 952

\bibitem[{{MacNeice} {et~al.}(2004){MacNeice}, {Antiochos}, {Phillips},
  {Spicer}, {DeVore}, \& {Olson}}]{MacNeice2004}
{MacNeice}, P., {Antiochos}, S.~K., {Phillips}, A., {et~al.} 2004, \apj, 614,
  1028

\bibitem[{{Masson} {et~al.}(2013){Masson}, {Antiochos}, \&
  {DeVore}}]{2013ApJ...771...82M}
{Masson}, S., {Antiochos}, S.~K., \& {DeVore}, C.~R. 2013, \apj, 771, 82

\bibitem[{{Mei} {et~al.}(2012){Mei}, {Shen}, {Wu}, {Lin}, {Murphy}, \&
  {Roussev}}]{2012MNRAS.425.2824M}
{Mei}, Z., {Shen}, C., {Wu}, N., {et~al.} 2012, \mnras, 425, 2824

\bibitem[{{Pariat} {et~al.}(2017){Pariat}, {Leake}, {Valori}, {Linton},
  {Zuccarello}, \& {Dalmasse}}]{Pariat2017}
{Pariat}, E., {Leake}, J.~E., {Valori}, G., {et~al.} 2017, \aap, 601, A125

\bibitem[{{Rice} \& {Yeates}(2022)}]{2022FrASS...9.9135R}
{Rice}, O. E.~K. \& {Yeates}, A.~R. 2022, Frontiers in Astronomy and Space
  Sciences, 9, 849135

\bibitem[{{Rice} \& {Yeates}(2023)}]{2023ApJ...955..114R}
{Rice}, O. E.~K. \& {Yeates}, A.~R. 2023, \apj, 955, 114

\bibitem[{{Sahade} {et~al.}(2022){Sahade}, {C{\'e}cere}, {Sieyra}, {Krause},
  {Cremades}, \& {Costa}}]{2022A&A...662A.113S}
{Sahade}, A., {C{\'e}cere}, M., {Sieyra}, M.~V., {et~al.} 2022, \aap, 662, A113

\bibitem[{{Sahade} {et~al.}(2023){Sahade}, {Vourlidas}, {Balmaceda}, \&
  {C{\'e}cere}}]{2023ApJ...953..150S}
{Sahade}, A., {Vourlidas}, A., {Balmaceda}, L.~A., \& {C{\'e}cere}, M. 2023,
  \apj, 953, 150

\bibitem[{{Talpeanu} {et~al.}(2022){Talpeanu}, {Poedts}, {D'Huys}, {Mierla}, \&
  {Richardson}}]{2022A&A...663A..32T}
{Talpeanu}, D.~C., {Poedts}, S., {D'Huys}, E., {Mierla}, M., \& {Richardson},
  I.~G. 2022, \aap, 663, A32

\bibitem[{{Titov} {et~al.}(2022){Titov}, {Downs}, {T{\"o}r{\"o}k}, \&
  {Linker}}]{Titov2022}
{Titov}, V.~S., {Downs}, C., {T{\"o}r{\"o}k}, T., \& {Linker}, J.~A. 2022,
  \apj, 936, 121

\bibitem[{{Wang} {et~al.}(2023){Wang}, {Liu}, \& {Luo}}]{2023AdSpR..72.5263W}
{Wang}, J., {Liu}, S., \& {Luo}, B. 2023, Advances in Space Research, 72, 5263

\bibitem[{{Wyper} {et~al.}(2021){Wyper}, {Antiochos}, {DeVore}, {Lynch},
  {Karpen}, \& {Kumar}}]{Wyper2021}
{Wyper}, P.~F., {Antiochos}, S.~K., {DeVore}, C.~R., {et~al.} 2021, \apj, 909,
  54

\bibitem[{{Yang} {et~al.}(2018){Yang}, {Dai}, {Chen}, {Li}, \&
  {Jiang}}]{2018ApJ...862...86Y}
{Yang}, J., {Dai}, J., {Chen}, H., {Li}, H., \& {Jiang}, Y. 2018, \apj, 862, 86

\bibitem[{{Zhang} {et~al.}(2020){Zhang}, {Wang}, {Liu}, {Zhang}, {Hu}, {Wang},
  {Zhuang}, \& {Li}}]{2020ApJ...898L..12Z}
{Zhang}, Q., {Wang}, Y., {Liu}, R., {et~al.} 2020, \apjl, 898, L12

\bibitem[{{Zhuang} {et~al.}(2018){Zhuang}, {Hu}, {Wang}, {Zhang}, {Liu}, {Gou},
  \& {Shen}}]{2018JGRA..123.2513Z}
{Zhuang}, B., {Hu}, Y., {Wang}, Y., {et~al.} 2018, Journal of Geophysical
  Research (Space Physics), 123, 2513

\end{thebibliography}

%\begin{thebibliography}{}
 
%\end{thebibliography}

\end{document}